\documentclass[runningheads,a4paper]{llncs}
\usepackage{times}
\usepackage{helvet}
\usepackage{courier}

\usepackage{subcaption}
\usepackage{float}
\usepackage{graphicx}
\usepackage{amsmath}
\usepackage{multirow}
\usepackage{fancyhdr}
\usepackage{booktabs}
\usepackage{paralist}
\usepackage{url}
\usepackage{color}
\usepackage{cite}
\usepackage{epstopdf}
\usepackage{tabularx}
\usepackage{algorithm}
\usepackage{algorithmic} 
\usepackage{changepage}
\newcolumntype{L}[1]{>{\raggedright\let\newline\\\arraybackslash\hspace{0pt}}m{#1}}
\newcolumntype{C}[1]{>{\centering\let\newline\\\arraybackslash\hspace{0pt}}m{#1}}
\newcolumntype{R}[1]{>{\raggedleft\let\newline\\\arraybackslash\hspace{0pt}}m{#1}}
\newcommand{\name}{}
\def\name/{SMLP}

\usepackage{url}
\urldef{\mailsa}\path|ahajibagheri@sonobi.com, gitars@eecs.ucf.edu|
\newcommand{\keywords}[1]{\par\addvspace\baselineskip
\noindent\keywordname\enspace\ignorespaces#1}

\begin{document}

\mainmatter  

\title{Investigating Negative Interactions in Multiplex Networks: A Mutual Information Approach}
\author{Alireza Hajibagheri\inst{1} \and Gita Sukthankar\inst{2}}

\authorrunning{Hajibagheri et al.}
\titlerunning{Investigating Negative Interactions in Multiplex Networks: A Mutual Information Approach}
%
%
%

\institute{
	\mbox{}\inst{1}Sonobi, Winter Park, Florida\\
	\inst{2}University of Central Florida, Orlando, Florida\\
\mailsa\\
}

%
%

\maketitle

\begin{abstract}
Many interesting real-world systems are represented as complex networks with multiple types of interactions and complicated dependency structures between layers.  These interactions can be encoded as having a valence with positive links marking interactions such as trust and friendship and negative links denoting distrust or hostility.  Extracting information from these negative interactions is challenging since standard topological metrics are often poor predictors of negative link formation, particularly across network layers.  In this paper, we introduce a method based on mutual information which enables us to predict both negative and positive relationships.  Our experiments show that \name/ (Signed Multiplex Link Prediction) can leverage negative relationship layers in multiplex networks to improve link prediction performance.

\end{abstract}

\section{Introduction}
\label{sec:intro}

While both positive and negative relationships clearly exist in many social network settings, the vast majority of research has only considered positive relationships. On social media platforms, people form links to indicate friendship, trust, or approval,  but they also link to signify disapproval of opinions or products. In this paper, we address the problem of predicting future user interactions from other layers of the network where a layer could represent either negative or positive user interactions. To do this, it is crucial to determine the correlation between layers.  Two network layers of opposite valence are likely to have negatively correlated link formation processes. To capture the interdependencies between different layers, we use mutual information to determine the sign of the correlation. Mutual information expresses the reduction in uncertainty due to another variable; we demonstrate that the average value of a layer's mutual information can be used to calculate the correlation of link formation processes across network layers with unknown valences. 

In online games it is even possible to directly attack other players' avatars or armies. The recent availability of signed networks in social media sites such as Epinions and Slashdot or online games such as Travian and EverQuest has motivated more research on signed network analysis~\cite{leskovec2010predicting,chiang2014prediction,kunegis2009slashdot}. This recent work has shown that negative links have significant added value over positive links for various analytical tasks. For example, a small number of negative links improves the performance of recommender systems in social media\cite{victor2009trust,ma2009learning, badami2018case, badami2017peeking, badami2017detecting}. Similarly, trust and distrust relations in Epinions can help users find high-quality and reliable reviews\cite{guha2004propagation}.

On the other hand, as social media platforms offer customers more interaction options, such as \textit{friending},  \textit{following}, and \textit{recommending}, analyzing the rich tapestry of interdependent user interactions becomes increasingly complicated. Although standard social network analysis techniques~\cite{scott2012social} offer useful insights about these communities, there is relatively little theory from the social sciences on how to integrate information from multiple types of online interactions. Hence, dynamic multiplex networks~\cite{kivela2014} offer a richer formalism for modeling the social fabric of online societies rather than organizing them into separate social networks representing the history of different forms of user interaction. A multiplex network is a multilayer network that shares the same set of vertices across all layers.

Link prediction algorithms~\cite{hajibagheri2016link,hajibagheri2016holistic2,hajibagheri2016holistic,liben2007link,menon2011link,scellato2011exploiting} have been implemented for many types of online social networks, including massively multiplayer online games and location-based social networks. Despite the fact that link prediction is a well studied problem, few link prediction techniques specifically address the problem of simultaneously predicting positive links across multiple networks~\cite{tang2012inferring,davis2013supervised,hristova2015multilayer,rossetti2011scalable}, and there is little literature available on predicting negative links in such networks~\cite{hajibagheri2017extracting,beigi2016signed,leskovec2010predicting}. 

The contributions of this paper can be summarized as follows:

\begin{compactitem}
\item  We investigate the role of mutual information in determining the sign of the correlation between different layers of a multiplex network.
\item  We introduce a framework for leveraging negative links in multiplex networks called SMLP (Signed Multiplex Link Prediction) that integrates mutual information to improve link prediction accuracy.
\item Extensive experiments are conducted on datasets collected from different types of social networks indicates that the proposed model outperforms state-of-the-art link prediction methods in cases where the networks contain user interactions with different valences.
\end{compactitem}

\noindent In the next section, we present related work on link prediction. The proposed framework is described in Section~\ref{sec:framework}. Section~\ref{sec:experiments} presents a detailed comparison between our our method and other well-known link prediction approaches. We conclude in Section~\ref{sec:conclusion} with a description of possible directions for future work.

\begin{figure}[!t]
  \centering 
\includegraphics[width=0.8\textwidth]{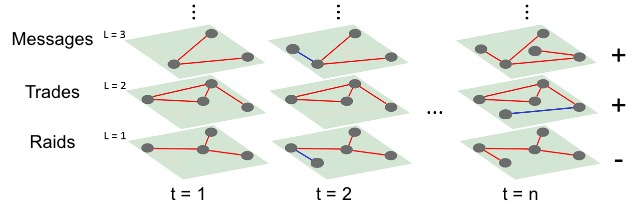}\label{sample_network}
    \caption{A sample complex network from an online game called Travian with both positive and negative correlation between layers. The figure represents three layers of the game network over time.}
    \label{fig:sample_network}
\end{figure}

\section{Related Work}
\label{sec:related}
A variety of computational approaches have been employed for predicting links in single layer networks, including supervised classifiers, statistical relational learning, matrix factorization, metric learning, and probabilistic graphical models (see surveys by \cite{lu2011link,al2011survey,zhang2014link,wang2015link} for a more comprehensive description). Regardless of the computational framework, topological network measures are commonly used as features to describe node pairs and can be combined in a supervised or unsupervised fashion to do link prediction~\cite{liben2007link}. In this paper, we aggregate several of these metrics (listed in the next section), but our framework can be easily generalized to include other types of features.

Recently several papers have begun to investigate negative as well as positive relationships in online contexts. For example, users on Epinions express trust or distrust of others\cite{guha2004propagation}, and Slashdot participants declare users to be either friends or foes~\cite{kunegis2009slashdot,lampe2007follow}. More generally, arbitrary hyperlinks on the Web can be used to indicate agreement or disagreement with the target of the link, though the lack of explicit labeling in this case makes it more difficult to reliably determine this sentiment\cite{pang2008opinion}. The sign of a given link in a social network is defined as positive or negative based on the attitude from the source of the link to the recipient. A fundamental question is then the following: How do negative links affect the formation of those with positive sign? Can we extract information from negative relationships to predict positive relationships and vice-versa? Answers to these questions can help us reason about how negative relationships are used in online systems. When used  correctly, negative link information benefits user searches. For example, the addition of a small number of negative links has been shown to improve the performance of recommender systems in social media\cite{victor2009trust,ma2009learning}. Similarly, trust and distrust relations in Epinions can help users find high-quality and reliable reviews\cite{guha2004propagation}.

The primary focus of this paper is leveraging correlations between layers to improve link prediction in multiplex networks. More specifically, we present a theoretical analysis of the link prediction problem from the perspective of information theory.  This process of using cross-layer information can be treated as a transfer learning problem where information is learned from a source network and applied to improve prediction performance the target network.  Tang et al.~\cite{tang2012inferring} introduced a transfer-based factor graph (TranFG) model which incorporates social theories into a semi supervised learning framework.  This model is then used to transfer supervised information from a source network to infer social ties in the target network.

Another strategy is to create more general versions of the topological measures that capture activity patterns in multilayer networks.  Davis et al.~\cite{davis2013supervised} introduced a probabilistically weighted extension of the Adamic/Adar measure for these networks. Weights are calculated by doing a triad census to estimate the probability of different link type combinations.  The extended Adamic/Adar metric is then used, along with other unsupervised link predictors, as input for a supervised classifier.  Similarly, Hristova et al.~\cite{hristova2015multilayer} extend the definition of network neighborhood by considering the union of neighbors across all layers. These multilayer features are then combined in a supervised model to do link prediction. One weakness with the above mentioned models is their inability to use temporal information accrued over many snapshots, rather than relying on a single previous snapshot.  In this paper, we evaluate two versions of our \name/ framework, a version that only uses topological metrics calculated from one time slice vs.\ multiple snapshots. Rossetti et al.~\cite{rossetti2011scalable} combined multidimensional versions of Common Neighbors and Adamic/Adar with predictors that are able to utilize temporal information.  However, like the standard version of these metrics, these extended versions do not necessarily generalize to networks generated from different processes.

Moreover, the process of information extraction from negative layers can be treated as measuring the average mutual information of a target layer based on a negative or positive predictor layer. Tan et al.\cite{tan2014link} developed a framework to uncover missing edges in networks via the mutual information of network topology. They estimate mutual information parameters for a node pair based on structural features of a network. Node pairs are then sorted according to their mutual information and those with highest values are deemed the best candidates to link in the future. Here, we use mutual information simply to determine the sign of correlations between layers; the scoring process is done separately.

\section{Proposed Method}
\label{sec:framework}

In previous work, we introduced MLP\cite{hajibagheri2016holistic}, a hybrid architecture that utilizes multiple components to address different aspects of the link prediction task. While MLP utilizes information from all layers of a network to improve link prediction in a target layer, it is not able to capture negative correlations among different layers of a network. For example, let us consider $\alpha$ and $\beta$ as layers of a network $N$ where $\alpha$ and $\beta$ represent positive (trades) and negative (raids) relationships between users in a game respectively. Now, if our goal is to predict future links of layer $\alpha$ using information from $\beta$, MLP fails to capture the negative effect of a link between two nodes on $\beta$. In order to model such relationships, we propose a model based on \textit{mutual information} that can capture the correlation sign between different layers in order to modify the MLP weighting procedure. In this section, we first provide a short description on the concept of mutual information and describe how it has been previously used for the link prediction task in single layer networks. Finally, we introduce our new signed multiplex link prediction model. 

MLP tries to extract information from all layers of the network for the purpose of predicting links within a specific layer known as the target layer. To do so, we create a weighted version of the original target layer where interactions and connections that exist in other layers receive higher weights. After reweighting the layer, we employ a collection of node similarity metrics on the weighted network. To express the temporal dynamics of the network, a decay model is used on the time series of similarity metrics to predict future values. Finally, the Borda rank aggregation method is employed to combine the ranked lists of node pairs into a single list that predicts links for the next snapshot of the target network layer. 

\subsection{Using Mutual Information for Link Prediction}

Considering a random variable $X$ associated with outcome $x_k$ with probability $p(x_k)$, its self-information $I(x_k)$ can be denoted as~\cite{shannon2001mathematical}:

\begin{equation}
I(x_k) = \log \frac{1}{p(x_k)} = -\log p(x_k)
\end{equation}

The higher the self-information is, the less likely the outcome $x_k$ occurs. On the other hand, the mutual information of two random variables can be denoted as:

\begin{equation}
I(x_k;y_j) = \log \frac{p(x|y)}{p(x)} = - \log p(x_k)-(-\log p(x_k|y_j)) = I(x_k) - I(x_k|y_j)
\end{equation}

The mutual information is the reduction in uncertainty due to another variable. Thus, it is a measure of the dependence between two variables. It is equal to zero if and only if two variables are independent. Tan et al.\cite{tan2014link} proposed the following link prediction model based on mutual information. Let $\Gamma(x)$ represent node $x$'s neighbors, then for the node pair $(x,y)$, the set of their common neighbors is denoted as:

\begin{equation}
O _ {xy} = \Gamma(x) \cap \Gamma(y)
\end{equation}

Given a disconnected node pair $(x,y)$, if the set of their common neighbors $O_{xy}$ is available, the likelihood score of node pair $(x,y)$ is defined as:

\begin{equation}
s_{xy}^{MI} = -I(L_{xy}^1 | O_{xy})
\end{equation}

\noindent where $I(L_{xy}^1 | O_{xy})$ is the conditional self-information of the existence of a link between node pair $(x,y)$ when their common neighbors are known (refer to\cite{tan2014link} for more details). According to the property of self-information, the smaller $I(L_{xy}^1 | O_{xy})$ is, the higher the likelihood of link existence, $I(L_{xy}^1 | O_{xy})$ can thus be derived as :

\begin{equation}
I(L_{xy}^1 | O_{xy}) = I(L_{xy}^1) - I(L_{xy}^1;O_{xy})
\end{equation}

\noindent where $I(L_{xy}^1)$ is the self-information of that node pair $(x,y)$ being connected and can be estimated as:

\begin{equation}
p(L_{xy}^1) = 1 - p(L_{xy}^0) = 1 - \frac{C_M^{k_x}-k_y}{C_M^{k_x}}
\end{equation}

\noindent where $M$ is the total number of links and $k_x$ is the degree of node $x$. 

Also, $I(L_{xy}^1;O_{xy})$ is the mutual information between the event that node pair $(x,y)$ is linked and the event that the node pair's common neighbors are known. Note that $I(L_{xy}^1)$ is calculated by the prior probability of node $x$ and node $y$ being connected. $I(L_{xy}^1;O_{xy})$ indicates the reduction in the uncertainty of nodes $x$ and $y$ linking due to the information provided by their common neighbors. Node pairs are then sorted based on their score $s_{xy}^{MI}$ to determine the best candidates for link formation in future timesteps. 

\subsection{Signed Multiplex Link Prediction}

\subsubsection{Multilayer Neighborhood}
\label{multilayer-neighborhood}
Various experiments on multilayer link prediction have indicated that using neighborhood information from different layers of a network in a multiplex environment can improve the performance of link prediction. Hristova et al.\cite{hristova2015multilayer} proposed the concept of a multilayer neighborhood where a link that exists on more than one layer in a multiplex network is called a \textit{multiplex link}. Following the definition of a multilayer network, the ego network of a node can be redefined as the multilayer neighborhood. While the simple node neighborhood is the collection of nodes one hop away from it, the multilayer global neighborhood (denoted by GN) of a node $i$ can be derived by the total number of unique neighbors across layers:

\begin{equation}
\Gamma_{GNi} = \{j \in V^{\mathcal{M}} : e_{i,j} \in E^{\alpha\cup\beta} \}
\end{equation}

\noindent Similarly, the core neighborhood (denoted by CN) of a node $i$ across layers of the multilayer network is defined as:

\begin{equation}
\Gamma_{CNi} = \{j \in V^{\mathcal{M}} : e_{i,j} \in E^{\alpha\cap\beta} \}
\end{equation}

\begin{figure}[!t]
  \centering 
\begin{subfigure}[t]{2.2in}
\includegraphics[width=1.0\textwidth]{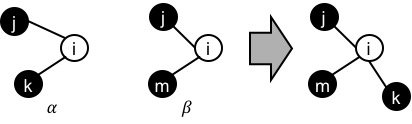}\label{mutual_others}
\caption{Global Neighborhood}\label{fig:1a}		
\end{subfigure}
\quad
\begin{subfigure}[t]{2in}
\includegraphics[width=1.0\textwidth]{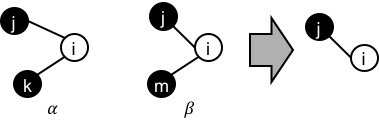}\label{mutual_others}
\caption{Core Neighborhood}\label{fig:1b}		
\end{subfigure}
\quad
    \caption{Sample networks representing global and core sets for node $i$ based on multilayer neighborhood definitions where that node exists on multiple layers of the network ($\alpha$ and $\beta$).}
    \label{fig:global-core}
\end{figure}

Figure~\ref{fig:global-core} shows an example of results for both multilayer neighborhood methods calculating neighbors of a node $i$. The goal is to determine the type of correlation (negative or positive) between two layers of a multiplex network. To this end, we calculate the average mutual information value for a target layer based on predictor layers. There are two assumptions: 
\begin{itemize}
\item Using the core neighborhood, if there is a negative correlation between two layers, the value of average mutual information would decrease substantially but would not change significantly if there is a positive correlation between the two layers.
\item If the global neighborhood is used to calculate the value of mutual information for a node pair, this would increase the value of average mutual information if there is a positive correlation between two layers and would not change significantly otherwise. 
\end{itemize}

After the sign of the relationship between the target layer and other predictor layers has been determined, MLP is used for predicting future links with the minor addition that layer weights can have both negative and positive values. 

\subsubsection{Mutual Information Definition} We now introduce the definitions of the self-information and of the mutual information, respectively.

\textbf{Definition 1\cite{shannon2001mathematical}} Considering a random variable $X$ associated with outcome $x_k$ with probability $p(x_k)$, its self-information $I(x_k)$ can be denoted as:

\begin{equation}
I(x_k) = \log \frac{1}{p(x_k)} = - \log p(x_k)
\end{equation}

where the base of the logarithm is specified as 2, thus the unit of self-information is bit. This is applicable for the following if not otherwise specified. The self-information indicates the uncertainty of the outcome $x_k$. Obviously, the higher the self-information is, the less likely the outcome $x_k$ occurs.

\textbf{Definition 2\cite{cover2012elements}} Consider two random variables $X$ and $Y$ with a joint probability mass function $p(x,y)$ and marginal probability mass functions $p(x)$ and $p(y)$. The mutual information $I(X; Y)$ can be denoted as follows:

\begin{equation}
\begin{split}
I(X; Y) &= \Sigma_{x \in X} \Sigma_{y \in Y} p(x,y) \log \frac{p(x,y)}{p(x)p(y)} \\ &= \Sigma_{x,y} p(x,y) \log \frac{p(x,y)}{p(x)p(y)} \\ &= \Sigma_{x,y} p(x,y) \log \frac{p(x|y)}{p(x)p(y)}
\end{split}
\end{equation}

Ultimately, the mutual information $I(x_k; y_j)$ could be defined as:

\begin{equation}
\begin{split}
I(x_k; y_j) &= \log \frac{p(x_k|y_j)}{p(x_k)p(y_j)} \\ 
& = -\log p(x_k) - (-\log p(x_k|y_j)) \\ 
& = I(x_k) - I(x_k|y_j)
\end{split}
\end{equation}

The mutual information is the reduction in uncertainty due to another variable. Thus, it is a measure of the dependence between two variables. It is equal to zero if and only if two variables are independent. Next we will describe how we are going to use mutual information for improved link prediction results.

\subsubsection{Proposed Framework} Assuming that we have a multilayer network where layers have the same set of nodes, given a disconnected node pair $(x, y)$, the mutual information of link existence (assuming that the set of their common neighbors $O_{xy}$ is available) can be derived as:

\begin{equation}
I(L_{xy}^1; O_{xy}) = I(L_{xy}^1) - I(L_{xy}^1 | O_{xy})
\end{equation}

\noindent where:

\begin{equation}
I(L_{xy}^1 | O_{xy}) = \sum_{z \in O_{xy}} I(L_{xy}^1 | z)
\end{equation}

For a multiplex network, $O_{xy}$ can be redefined to use the global and core neighborhood of the two nodes as described in\ref{multilayer-neighborhood}. As a result, the average mutual information of a target layer would be defined as:

\begin{equation}
MI^{\alpha} = \frac{1}{|E^{\alpha}|} \sum_{x,y \in E^{\alpha} \& x \neq y} I(L_{xy}^1; O_{xy})
\end{equation}

The value of $MI^{\alpha}$ is calculated using information from a predictor layer $\beta$. It can be used to indicate the sign (negative or positive) of the correlation between the link formation in two layers. This sign is then  used within the second phase of link prediction task (MLP) to assign weights to all predictor layers and hence improve the performance of link prediction task for the target layer $\alpha$. More details on the relationship between the value of $MI^{\alpha}$ and the sign of correlation between layers are provided in Section~\ref{subsec:analysis}.

\section{Experimental Study}
\label{sec:experiments}

This paper evaluates the \name/ framework on networks extracted from two real-world datasets, Travian and Cannes2013.  Not only do we compare our results with two other approaches for fusing cross-layer information, but we also consider scores generated by mutual information paired with core and global neighborhood as distinct methods. This enables us to investigate the impact of each multiplex neighborhood metric on the final results. Hence, \name/ is compared with our previous model MLP, Mutual Information (N) a  single layer link prediction method, Mutual Information (CN) and (GN) which incorporate core and global neighborhood definitions respectively. We also incorporate multiple unsupervised methods such as Common Neighbors and Adamic/Adar as baseline. For the sake of fair comparison and due to the reason that SMLP incorporates weighted network, we use the weighted version of the unsupervised methods as well. All of the algorithms were implemented in Python and executed on a machine with the Intel(R) Core i7 CPU and 24GB of RAM for the purpose of fair comparison. The code\footnote{https://github.com/alirezahajibagheri/LinkPredictionPackage} and related datasets could be found on our group webpage\footnote{http://ial.eecs.ucf.edu/travian.php}. 

\subsection{Datasets}
We use two real-world dynamic multiplex networks to demonstrate the performance of our proposed algorithm. These networks are considerably disparate in structure and were selected from different domains. Negative links (raids) exist between users of our MMOG dataset to evaluate the performance of our method in predicting negative links as well as examining these layers on positive ones. Table~\ref{tab:network_stats} provides the network statistics for each of the datasets:
\begin{itemize}
\setlength\itemsep{1em}
\item \textbf{Travian MMOG~\cite{hajibagheri2015conflict}} 
Travian is a popular browser-based real-time strategy game with more than 5 million users.  Players seek to improve their production capacity and construct military units in order to expand their territory through a combination of colonization and conquest. Each game cycle lasts a fixed period (a few months) during which time the players vie to complete construction on one of the Wonders of the World. To do this, they form alliances of up to 60 members under a leader or a leadership team; in this article these alliances are used as the ground truth for evaluating the community detection procedure.  

Travian has an in-game messaging system (IGM) for player communication which was used to create our Messages network. Each player can submit a request to trade a specific resource. If another player finds this request interesting, he/she can accept it and the trade will occur; this data was used to build the Trade network. In this research, we have used three months of data collected from a Travian server in Germany which was specifically set up for research purposes. We focus on networks created from trades and messages as well as raids which represents a negative relationship between players and hence could have a negative correlation with other layers of the network\cite{hajibagheriusing}. 

\item \textbf{Twitter Interactions}~\cite{omodei2015characterizing} This dataset consists of Twitter activity before, during, and after an ``exceptional'' event as characterized by the volume of communications.  Unlike most Twitter datasets which are built from follower-followee relationships, links in this multiplex network correspond to retweeting, mentioning, and replying to other users. The Cannes2013 dataset was created from tweets about the Cannes film festival that occurred between May 6,2013 to June 3, 2013. Each day is treated as a separate network snapshot.
\end{itemize}

\begin{table}[t]
\renewcommand{\arraystretch}{1.2}
\caption{Dataset Summary (Network Layers)}
\label{tab:network_stats}
\begin{center}
\begin{tabular}{llclccc}
\toprule
\textbf{Dataset} & & 
\textbf{Travian} & & 
\textbf{Cannes2013} & &
 \\
\midrule
\textbf{No. of Nodes} & & 2,809 & &438,537& \\
\textbf{No. of Snapshots} & & 30 & & 29 &
\\
\midrule
\textbf{Layers/No. of Edges} 
\\
\midrule
\textbf{No. of nodes} & & 150 & & 2,809\\
\textbf{Link (Class 1)} & & 5,015 & & 44,956\\
\textbf{No Link (Class 0)} & & 17,485 & & 7,845,5256\\
& Trades & 87,418 & Retweet & 496,982 & & \\
& Messages & 44,956 & Mention & 411,338 & & \\
& Raids & 56,765 & Reply & 83,534 & & \\

\bottomrule
\end{tabular}
\end{center}
\end{table}

\subsection{Evaluation Metrics}
For the evaluation, we measure receiver operating characteristic (ROC)  curves for the different approaches. The ROC curve is a plot of the \textit{true positive rate (TPR)} against the \textit{false positive rate (FPR)}. We report area under the ROC curve (AUROC), the scalar measure of the performance over all thresholds. 

These curves show achievable true positive rates (TP) with respect to all false positive rates (FP) by varying the decision threshold on probability estimations or scores. For all of our experiments, we report area under the ROC curve (AUROC), the scalar measure of the performance over all thresholds. Since link prediction is highly imbalanced, straightforward accuracy measures are well known to be misleading; for example, in a sparse network, the trivial classifier that labels all samples as missing links can have a 99.99\% accuracy.

\subsection{Baselines}
\label{subsec:baselines}

To assess our proposed framework and study the impact of its components, we compare against the following baselines:

\begin{itemize}
\setlength\itemsep{1em}
\item \textbf{Multiplex Link Prediction Model (MLP)}:  This model utilizes the likelihood assignment and edge weighting procedure to extract cross-layer information from a network. Finally, node similarity scores are modified using the temporal decay model and combined with Borda rank aggregation to generate scores for unconnected node pairs.\cite{hajibagheri2016holistic2} 

\item \textbf{Number of Common Neighbors (CN)}
The CN measure is defined as the number of nodes with direct relationships with both evaluated nodes $x$ and $y$~\cite{Newman01clusteringand}. For weighted networks, the CN measure is:

\begin{equation}\label{eq:CNW}	CN(x,y)=\sum_{z\in|\Gamma(x)\cap\Gamma(y)|}w(x,z)+w(y,z)
\end{equation}

\item \textbf{Jaccard's Coefficient (JC)}

The JC measure assumes higher values for pairs of nodes who share a higher proportion of common neighbors relative to their total neighbors.

\begin{equation}\label{eq:JCW}
	JC(x,y) = \frac{\sum_{z \in \Gamma(x)\cap\Gamma(y)} w(x,z) + w(y,z)}{\sum_{a\in\Gamma(x)} w(x,a) + \sum_{b\in\Gamma(y)} w(y,b)}
\end{equation}

\item \textbf{Preferential Attachment (PA)}

The PA measure assumes that the probability that a new link originates from  node $x$ is proportional to its node degree. Consequently, nodes that already possess a high number of relationships tend to create more links ~\cite{barabasi2009scale}.

\begin{equation}\label{eq:PAW}
	PA(x,y)=\sum_{z_{1}\in\Gamma(x)} w(x,z_1) \times \sum_{z_{2}\in\Gamma(y)} w(y,z_2)
\end{equation}

\item \textbf{Adamic-Adar Coefficient (AA)}

This metric~\cite{adamic2003friends} is closely related to Jaccard's coefficient in that it assigns a greater importance to common neighbors who have fewer neighbors. Hence, it measures the exclusivity of the relationship between a common neighbor and the evaluated pair of nodes.

\begin{equation}\label{eq:AAW}	AA(x,y)=\sum_{z\in\Gamma(x)\cap\Gamma(y)}\frac{w(x,z) + w(y,z)}{log(1 + \sum_{c\in\Gamma(z)} w(z,c))}
\end{equation}

\item \textbf{Resource Allocation (RA)}
	
RA was first proposed in~\cite{zhou2009predicting} and is based on physical processes of resource allocation.

\begin{equation}\label{eq:RAW}	RA(x,y)=\sum_{z\in\Gamma(x)\cap\Gamma(y)}\frac{w(x,z) + w(y,z)}{\sum_{c\in\Gamma(z)} w(z,c)}
\end{equation}

\item \textbf{Page Rank (PR)} 
	
The PageRank algorithm~\cite{brin2012reprint} measures the significance of a node based on the significance of its neighbors.  We use the weighted PageRank algorithm proposed in\cite{ding2011applying}.

\begin{equation}\label{eq:PRW}
	PR_{w}(x)= \alpha \sum_{k\in\Gamma(x)} \frac{PR_{w}(x)}{L(k)} + (1-\alpha) \frac{w(x)}{\sum_{y=1}^{N} w(y)}
\end{equation}

where $L(x)$ is the sum of outgoing link weights from node $x$, and $\sum_{y=1}^{N} w(y)$ is the total weight across the whole network.

\item \textbf{Inverse Path Distance (IPD)}

The Path Distance measure for unweighted networks simply counts the number of nodes along the shortest path between $x$ and $y$ in the graph. Note that $PD(x,y) = 1$ if two nodes $x$ and $y$ share at least one common neighbor. In this article, the Inverse Path Distance is used to measure the proximity between two nodes, where:

\begin{equation}\label{eq:IPDW}
IPD(x,y) = \frac{1}{PD(x,y)}
\end{equation}

IPD is based on the intuition that nearby nodes are likely to be connected. In a weighted network, IPD is defined by the inverse of the shortest weighted distance between two nodes. 

\item \textbf{Product of Clustering Coefficient (PCF)} 

The clustering coefficient of a vertex $v$ is defined as:

\begin{equation}
	PCF(v) = \frac{3 \times \mbox{\# of triangles adjacent to v}}{\mbox{\# of possible triples adjacent to v}}
\end{equation}
To compute a score for link prediction between the vertex $x$ and $y$, one can multiply the clustering coefficient score of $x$ and $y$.

\item \textbf{Average Aggregation}: In order to extend the rank aggregation model to include information from other layers of the network, we use the idea proposed in~\cite{pujari2015link}. Node similarity metrics are aggregated across all layers. So for attribute $X$ (Common Neighbors, Adamic/Adar, etc.) over $M$ layers the following is defined:

\begin{equation}
X(u,v) = \frac{\sum_{\alpha=1}^M X(u,v)^{\alpha}}{M}
\end{equation}

where $X(u,v)$ is the average score for nodes $u$ and $v$ across all layers and $X(u,v)^{\alpha}$ is the score at layer $\alpha$.  Borda's rank aggregation is then applied to the extended attributes to calculate the final scoring matrix.
\item \textbf{Entropy Aggregation}: Entropy aggregation is another extended rank aggregation model proposed in~\cite{pujari2015link} where $X(u,v)$ is defined as follows:

\begin{equation}
X(u,v) = -\sum_{\alpha=1}^M \frac{X(u,v)^{\alpha}}{X_{total}} \log(\frac{X(u,v)^{\alpha}}{X_{total}})
\end{equation}

where 

\begin{equation}
X_{total} = \sum_{\alpha=1}^M X(u,v)^{\alpha}
\end{equation}

The entropy based attributes are more suitable for capturing the distribution of the attribute value over all dimensions~\cite{nouraniselection,nourani2014using}.A higher value indicates a uniform distribution of attribute values across the multiplex layers. 
\item \textbf{Multiplex Common Neighbors}: Finally, using the definition of core neighborhood proposed in~\cite{hristova2015multilayer}, we extend four unsupervised methods (Common Neighbors, Preferential Attachment, Jaccard Coefficient and Adamic/Adar) to their multiplex versions.
\end{itemize}

\begin{figure}[!t]
  \centering 
 \begin{subfigure}[t]{2.5in}
\includegraphics[width=1.0\textwidth]{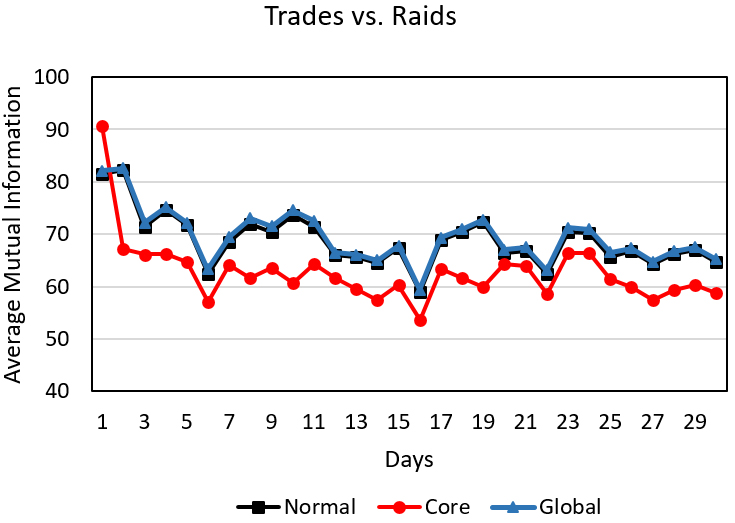}\label{travian1}\caption{}
\end{subfigure}
\begin{subfigure}[t]{2.5in}
\includegraphics[width=1.0\textwidth]{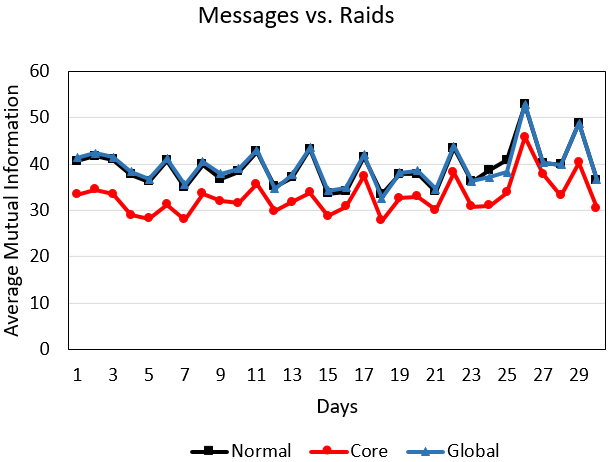}\label{travian2}\caption{}
\end{subfigure}
\begin{subfigure}[t]{2.5in}
\includegraphics[width=1.0\textwidth]{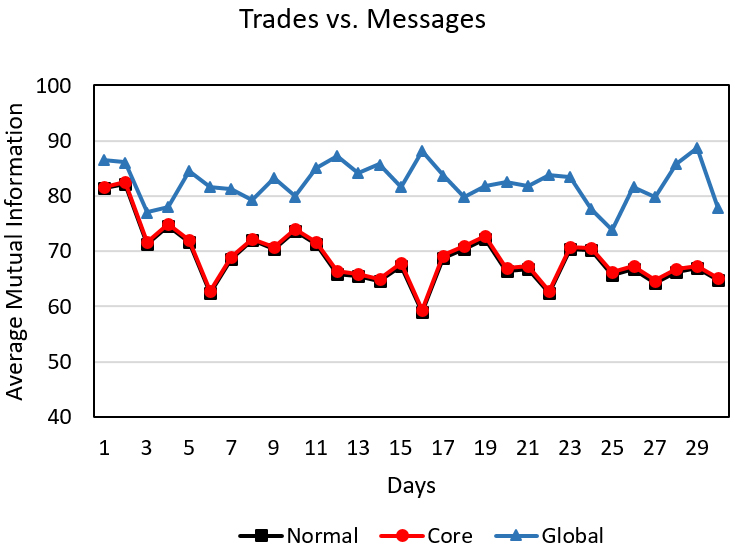}\label{travian3}\caption{} 
\end{subfigure}
    \caption{Average mutual information values over time calculated using normal and core neighborhood definitions.}
    \label{fig:avg_mi}
\end{figure}

\subsection{Analysis of Multilayer Neighborhood}
\label{subsec:analysis}

As mentioned before, our assumption is that both core and global neighborhood definitions would enable us to study correlations between different layers of a network. Figure~\ref{fig:avg_mi} shows average mutual information values for Travian network for each day of the 30 day period. There are negative and positive correlations between trades (or messages)-raids and trades-messages respectively. As shown in the figure, at every timestep of the network, the average value decreases when the core neighborhood is used in the case of negative correlation (Figure\ref{fig:avg_mi} (a) and (b)) and does not change significantly when global neighborhood is used to calculate the average value. On the other hand, as shown in Figure~\ref{fig:avg_mi} (c), when dealing with positive correlations (trades-messages), the value of average mutual information does not change significantly using the core neighborhood but increases drastically using the global neighborhood definition (the average difference is less than one). Hence, this justifies our assumption that the average mutual information value of a certain layer can be used to determine the sign of its role in target layer link prediction.

\begin{figure}[!t]
  \centering 
\includegraphics[width=0.8\textwidth]{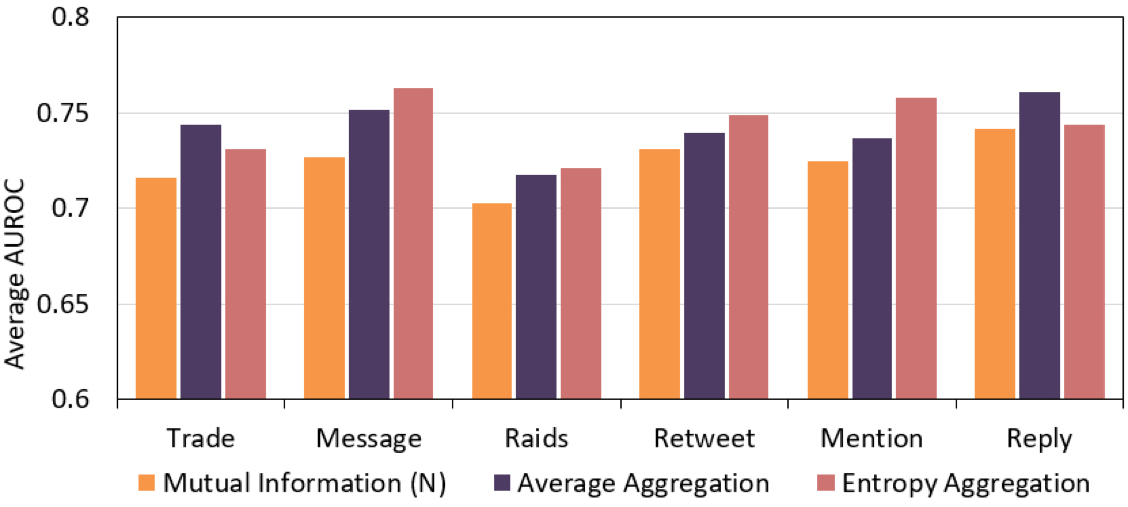}\label{mutual_others}
    \caption{Comparison between the Normal Mutual Information method (Mutual Information (N)) vs. Average Aggregation and Entropy Aggregation methods. The last two methods use aggregated information from all layers of a network and hence are able to outperform the normal mutual information method which uses no extra information from any other layer.}
    \label{fig:mi_others}
\end{figure}

\begin{figure}[!t]
  \centering 
\includegraphics[width=0.8\textwidth]{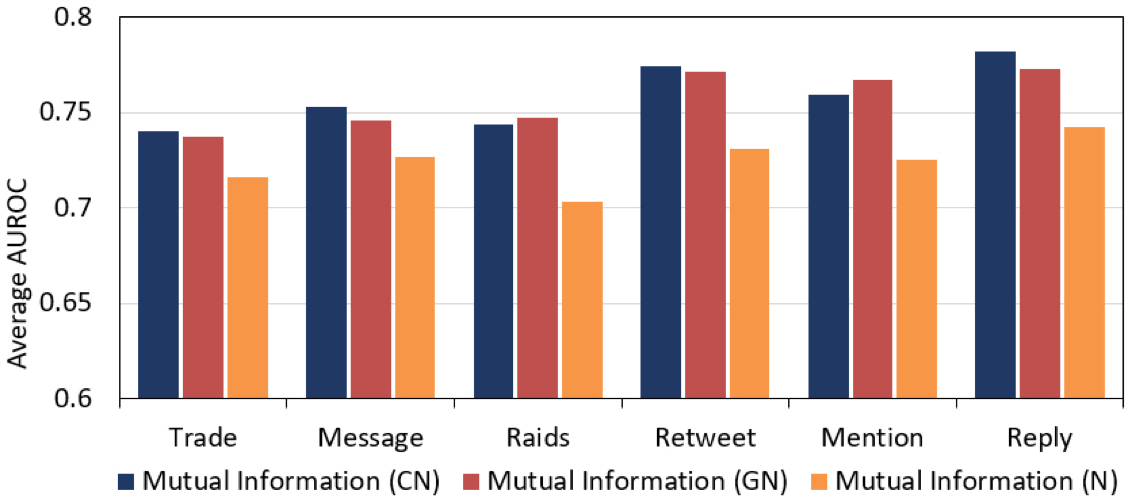}\label{all-mutuals}
    \caption{Comparison between mutual information based link prediction methods. Both methods that are paired with neighborhood methods outperform normal mutual information where core neighborhood is getting the best AUROC score. This is due to the same reason as before: incorporating information from other layers of the network.}
    \label{fig:mi_mutuals}
\end{figure}

\begin{figure}[!t]
  \centering 
\includegraphics[width=0.8\textwidth]{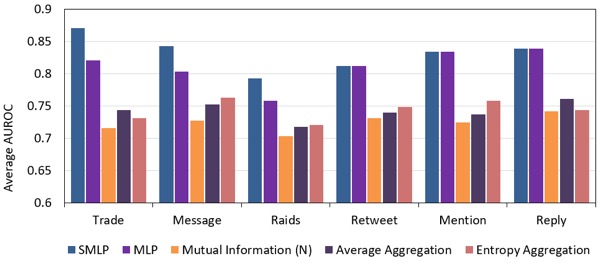}\label{all-methods}
    \caption{Signed Multiplex Link Prediction (SMLP) performance compared with MLP, normal mutual information and two aggregation methods. SMLP is able to outperform all since it uses cross-layer information from both positive and negative layers.}
    \label{fig:mi_all}
\end{figure}

\subsection{Performance of Signed Multiplex Link Prediction (SMLP)} 

Table~\ref{tab:auroc} shows the results of different algorithms on the Travian and Cannes2013 datasets. With 30 days of data from Travian and 27 days for Cannes2013, we were able to extensively compare the performance of the proposed methods and the impact of using different elements. AUROC performances for a target layer averaged over all snapshots are calculated, and our proposed framework is shown at the top of the table, followed by variants of mutual information based link prediction models using different definitions of neighborhood (N which stands for Normal proposed by Tan et al.\cite{tan2014link}, CN stands for Core Neighborhood and GN stands for Global Neighborhood). The algorithms shown in the bottom half of the table (Average Aggregation and Entropy Aggregation\cite{pujari2015link}) are techniques for multiplex networks proposed by other research groups. The settings given in\cite{hajibagheri2016holistic} were used for MLP. 

\begin{table}[!t]
\begin{adjustwidth}{-0.1\textwidth}{-0.1\textwidth}
\caption{AUROC performances for a target layer averaged over all snapshots and ten runs.}
	
\label{tab:auroc}
\begin{center}
  \bgroup
  \def\arraystretch{1.3}
  \begin{tabular*}{1.2\textwidth}{p{3.5cm}C{1.9cm}ccccc}
  	\hline
	Algorithms / Networks & Trade & Message & Raids & Retweet & Mention & Reply\\ \hline
\textbf{\name/} &\textbf{0.871$\pm$0.013}&\textbf{0.843$\pm$0.031}&\textbf{0.793$\pm$0.007}&\textbf{0.812$\pm$0.002}&\textbf{0.834$\pm$0.003}&\textbf{0.839$\pm$0.002} \\
\textbf{MLP} &0.821$\pm$0.001&0.803$\pm$0.002&0.758$\pm$0.001&0.812$\pm$0.002&0.834$\pm$0.003&0.839$\pm$0.002 \\
\textbf{Mutual Information (CN)} &0.740$\pm$0.016&0.753$\pm$0.011&0.744$\pm$0.013&0.774$\pm$0.009&0.759$\pm$0.012&0.782$\pm$0.005\\
\textbf{Mutual Information (GN)} &0.737$\pm$0.010&0.746$\pm$0.011&0.747$\pm$0.009&0.771$\pm$0.011&0.767$\pm$0.012&0.773$\pm$0.009\\
\textbf{Mutual Information (N)} &0.716$\pm$0.012&0.727$\pm$0.006&0.703$\pm$0.012&0.731$\pm$0.007&0.725$\pm$0.013&0.742$\pm$0.014\\
    \hline
\textbf{Common Neighbors} &0.656$\pm$0.002&0.667$\pm$0.002&0.699$\pm$0.002&0.705$\pm$0.003&0.699$\pm$0.001&0.712$\pm$0.031\\
\textbf{Jaccard Coefficient} &0.628$\pm$0.002&0.680$\pm$0.003&0.594$\pm$0.002&0.733$\pm$0.002&0.711$\pm$0.003&0.688$\pm$0.013\\
\textbf{Preferential Attachment} &0.709$\pm$0.002&0.637$\pm$0.001&0.584$\pm$0.002&0.612$\pm$0.002&0.587$\pm$0.003&0.601$\pm$0.003\\
\textbf{Adamic/Adar} &0.635$\pm$0.003&0.700$\pm$0.003&0.700$\pm$0.002&0.642$\pm$0.002& 0.516$\pm$0.003&0.630$\pm$0.002\\
\textbf{Resource Allocation} &0.625$\pm$0.005&0.690$\pm$0.003&0.597$\pm$0.002&0.622$\pm$0.002&0.672$\pm$0.003&0.693$\pm$0.003\\
\textbf{Page Rank} &0.595$\pm$0.0016&0.687$\pm$0.002&0.660$\pm$0.002&0.630$\pm$0.003&0.613$\pm$0.002&0.610$\pm$0.002\\
\textbf{Inverse Path Distance} &0.572$\pm$0.003&0.650$\pm$0.003&0.631$\pm$0.003&0.641$\pm$0.002&0.561$\pm$0.004&0.553$\pm$0.005\\
\textbf{Clustering Coefficient} &0.580$\pm$0.002&0.633$\pm$0.003&0.570$\pm$0.020&0.621$\pm$0.011&0.522$\pm$0.004&0.547$\pm$0.003\\ 
\hline
\textbf{Average Aggregation} &0.744$\pm$0.030&0.752$\pm$0.020&0.658$\pm$0.017&0.740$\pm$0.003&0.737$\pm$0.011&0.761$\pm$0.003\\
\textbf{Entropy Aggregation} &0.731$\pm$0.004&0.763$\pm$0.020&0.661$\pm$0.009&0.749$\pm$0.0030&0.758$\pm$0.031&0.744$\pm$0.002\\ 
\textbf{Multiplex CN} &0.729$\pm$0.0040&0.643$\pm$0.013&0.672$\pm$0.003&0.716$\pm$0.003&0.733$\pm$0.002&0.693$\pm$0.022\\
\textbf{Multiplex JC} 
&0.666$\pm$0.031&0.619$\pm$0.012&0.580$\pm$0.003&0.736$\pm$0.002&0.722$\pm$0.002&0.712$\pm$0.012\\
\textbf{Multiplex PA} 
&0.722$\pm$0.010&0.646$\pm$0.012&0.580$\pm$0.003&0.640$\pm$0.003&0.621$\pm$0.003&0.640$\pm$0.020\\
\textbf{Multiplex AA} 
&0.671$\pm$0.010&0.690$\pm$0.031&0.671$\pm$0.003&0.669$\pm$0.003& 0.552$\pm$0.003 & 0.612$\pm$0.013\\
    \hline
  \end{tabular*}
  \egroup
 	\end{center}
    \end{adjustwidth}
\end{table}

Bold numbers indicate the best results on each target layer considered. As expected, \name/ is the best performing algorithm in all cases since not only it utilizes both historical and cross-layer information, but also mutual information enables \name/ to capture negative correlations between different layers. As a result, node pairs that are connected by raids in Travian, are penalized for this connection and eventually receive a lower score compared to node pairs that are only connected on messages and trades. This holds true when raids is the target layer and two nodes are connected on either messages or trades layers which are negatively correlated with raids. On the other hand, it is evident that methods specifically designed for multiplex link prediction outperform  Mutual Information (N) which is unable to leverage cross-layer information. Also, Average Aggregation and Entropy Aggregation are able to achieve higher AUROC scores compared with Mutual Information based methods since they collect more information from the network using different similarity metrics such as common neighbors, Adamic/Adar, etc. Finally, for Twitter layers, \name/ and MLP achieve similar results since there are no negative layers to modify the sign of the weights associated with different layers of the network. Figures \ref{fig:mi_others}, \ref{fig:mi_mutuals} and \ref{fig:mi_all} provide detailed analysis of all models performance results on different datasets.

\section{Conclusion and Future Work}
\label{sec:conclusion}
In this paper, we introduce a new link prediction framework, \name/ (Signed Multiplex Link Prediction), that employs a holistic approach to accurately predict links in dynamic multiplex networks by incorporating negative relationships between users. Our analysis on real-world networks created by a variety of social processes suggests that \name/ effectively models multiplex network coevolution in many domains and is also able to capture negative correlation between layers in order to improve link prediction performance.

In future work, we are planning to extend our results to other multilayer networks that contain negative interactions; we are particularly interested in studying datasets that contain multiple negative interaction layers to see if they result in positive correlations. Another promising line of inquiry is studying the usage of multiple features (beyond common neighbors) to calculate mutual information, since different structural features highlight different aspects of network formation.

\section*{Acknowledgments}
The Travian dataset was provided by Drs.\ Rolf T. Wigand and Nitin Agarwal (University of Arkansas at Little Rock, Department of Information Science); their research was supported by the National Science Foundation and Travian Games GmbH, Munich, Germany.

\bibliographystyle{splncs}
\bibliography{references}
\end{document}